\documentstyle{article}
\newcommand{\dir}{.}
\begin{document}
\large
\baselineskip=0.8cm
\input psfig

\title{Diblock Copolymers at a\\ Homopolymer--Homopolymer--Interface:\\ A Monte Carlo Simulation}
\author{A. Werner, F. Schmid, K. Binder \\
Johannes Gutenberg Universit\"at Mainz\\ D--55099 Mainz \\
and \\
M. M\"uller\\
University of Washington\\ Department of physics, Box 351560\\ 
Seattle, Washington 98195}

\date{\today}
\maketitle
\begin{abstract}
The properties of diluted symmetric $A$-$B$ diblock copolymers 
at the interface between $A$ and $B$ homopolymer phases are studied by means 
of Monte Carlo (MC) simulations of the bond fluctuation model. We calculate 
segment density profiles as well as orientational properties of segments,
of $A$ and $B$ blocks, and of the whole chain. Our data support
the picture of oriented ``dumbbells'', which consist of mildly perturbed 
$A$ and $B$ Gaussian coils. The results are compared to a self consistent 
field theory (SCFT) for single copolymer chains at a homopolymer interface.
We also discuss the number of interaction contacts between monomers,
which provide a measure for the ``active surface'' of copolymers or
homopolymers close to the interface.
\end{abstract}
\newpage

\section{Introduction}

Blending of polymeric substances is a straightforward and inexpensive
way of creating new materials with improved mechanical properties\cite{mixt}.
However, long polymers of different type $A$ and $B$ are often immiscible 
already at high temperatures, since the total energy of the (usually repulsive) 
relative interaction is proportional to the total number of monomers and 
cannot be balanced by the entropy of mixing, which is proportional to the 
number of polymers in the mixture\cite{degennes,polint}. 
In order to overcome this problem, block copolymers containing
both types of monomers can be used as effective compatibilizers\cite{cop,B4}. 
Being partly compatible with both the $A$ and $B$ rich phase, they tend to
aggregate at interfaces, where they reduce the number of direct contacts
between $A$ and $B$ homopolymers, thereby reducing the interfacial
tension\cite{AGK}. Consequently the total area of interfaces increases, 
and the immiscible components may get finely dispersed in the mixture. 
Furthermore, copolymers improve the mechanical properties
of such interfaces: Due to entanglement between homopolymers and
copolymers, they increase the adhesive attraction and the fracture
toughness\cite{BR,Dai1,LC}. At high enough copolymer concentrations, additional
copolymer rich phases emerge which may display a diversity of
structures ordered on a mesoscopic scale\cite{phdexp,phdth}.

The simplest possible copolymers are diblocks, which consist of a block of
$A$ monomers connected to a block of $B$ monomers. From a thermodynamic
point of view, their effect on interfaces can be described as follows:
Copolymers act as amphiphiles in the homopolymer mixture\cite{HS}. Increasing 
the copolymer concentration causes the interfacial tension to decrease
monotonically, until this process is terminated by the formation of a
third, copolymer rich phase, e.g. an ordered lamellar phase or
microemulsion. The amphiphilic strength of a copolymer, {\em i.e.}, the 
maximum reduction
of the interfacial tension it can achieve, increases with the length of
the $A$ and $B$ blocks relative to the size of the homopolymers. Mean field
theories predict that the interfacial tension can be driven to zero
for long copolymers, which implies that the copolymer rich lamellar phase
evolves into the two-phase region via a continuous unbinding 
transition\cite{MWM,RI}. Fluctuations push the transition to first order in
real systems\cite{B}. One can argue that the low interfacial tensions at the 
presence of copolymers result from the affinity of the system to such 
an unbinding transition\cite{MS}.

This macroscopic discussion however does not shed light on the 
microscopic mechanisms, why and how copolymers work as compatibilizer or 
amphiphiles. In a simple microscopic picture, the $A$ block of copolymers 
at interfaces preferably stick into the $A$-rich phase, the $B$ block into 
the $B$-rich phase, and the molecules as a whole act as reinforcing 
rods\cite{BR}. The conformations of the copolymers determine the properties 
of the interface. 
A more detailed microscopic description of copolymer properties at interfaces 
has been developed by Leibler\cite{L2} and refined by Semenov\cite{SE}.
The analysis is based on the assumption, that the junction points between
$A$ and $B$ blocks are confined to a narrow region, which is much smaller
than the width of the copolymer layer at the interface. The presence of
copolymers at the interface gives rise to two free energy 
contributions: the entropy of mixing, divided into a translational
term and a swelling term, and the elastic energy of stretching of the
copolymer blocks. A scenario emerges which distinguishes between 
four different regimes: In the dilute regime, copolymers aggregate
at the interface, but do not yet overlap. The $A$ and $B$ blocks are
described as weakly perturbed coils, and the free energy is dominated by 
the mixing energy. When the mean interchain distance gets of the order of
the block radii of gyration, copolymers start to stretch and form a 
``wet brush''. At even higher copolymer concentrations, the ``dry brush'' 
regime is entered, where homopolymers do not penetrate into the interfacial 
region; $A$ and $B$ blocks have the conformation of stretched coils
and the free energy is dominated by the elastic energy of stretching.
Finally, in the saturated regime, the density of copolymers at the
interface is close to one and the interfacial segregation of copolymers
competes with the formation of micelles in the bulk\cite{SE}.

Experimental studies of copolymers at homopolymer interfaces have
been carried out by numerous groups, mostly using neutron reflectivity
or forward recoil spectroscopy \cite{LH}-\cite{SMR}. By deuterating individual
parts of the homopolymers or the copolymers, the excess of copolymers
at the interface can be measured as well as distributions of
homopolymer segments, copolymer segments or even selected copolymer
segments, like the junction point between $A$ and $B$ or the
end segment\cite{SMR,MA1,MA2}. These studies have provided detailed
insight into the microscopic structure of such interfaces.

Simple Leibler type theories are already quite successful in reproducing 
many of the experimental results\cite{SE,GR}. In order to reach quantitative
agreement, however, the Flory Huggins parameter $\chi$ has to be treated as 
adjustable, molecular weight dependent parameter. Less transparent, 
but more accurate mean field approaches are the self consistent
field theories\cite{HF1,HF2} or the density functional theories\cite{FR1,FR2}.
The self consistent field theory has first been applied to copolymer-homopolymer 
interfaces by Noolandi and Hong\cite{NH} and adjusted
to the case of a blend without solvent by Shull and Kramer\cite{SK,S}.
It has been shown to be quantitatively successful in predicting
the correct copolymer excess at the interface, using a $\chi$-parameter 
which is taken from independent bulk measurements\cite{SMR}. The
calculated width of the segment interface is somewhat too low, but
the discrepancies can be understood quantitatively if broadening due
to capillary waves is accounted for\cite{Dai2,SMR}. 
Such a remarkable success of a mean field theory is characteristic for 
polymeric substances, in which high molecular weight 
polymers interact with a high number of other polymers\cite{degennes,KB1}. 
In addition to reproducing experimental data, the self consistent
field theory also has the advantage of yielding further structural 
information, e.g. on chain conformations, monomer orientations 
etc.\cite{FT,SM}, which may be hard to access experimentally. Unfortunately, 
the theory also has some serious drawbacks. In particular, the usual 
treatment of polymers as Gaussian random walks is questionable for chains of 
the minority component, e.g. $A$ in the $B$ rich phase, and 
generally on length scales smaller than the screening length of the 
excluded volume\cite{SM}.

Computer simulations provide another way of obtaining additional information
on the microscopic structure of interfaces. Simulations
of inhomogeneous polymeric systems are computationally extremely
demanding, and therefore rare. Minchau et al~\cite{MDB} and 
Fried and Binder\cite{FB,BF} have investigated the phase behavior of
pure copolymer systems. Pan et al studied microphase structures in systems 
of short copolymers, which are swollen by a small amount 
(volume fraction 10 \%) of longer homopolymers\cite{PHB}.
Wang and Mattice have studied the adsorption of self avoiding copolymers at 
a stationary, sharp interface, modelled by an external field with a sharp 
kink\cite{WM}. A similar study has been performed for a random copolymer by 
Peng et al~\cite{PSB}. 
More detailed simulations have been presented by Cifra\cite{cifra}.

In this work, we present a study of copolymers at the interface
between homopolymer phases, where both homopolymers 
and copolymers are treated in microscopic detail. 
Pure homopolymer interfaces in immiscible 
$A$/$B$ blends have been analyzed previously\cite{MBO} and compared 
to the predictions of a self consistent field theory\cite{SM}. 
Here, we consider the effect of adding a small number 
of symmetric diblock copolymers to such a ``known'' homopolymer
interface. We restrict ourselves to the diluted case, where the copolymer 
coils do almost not overlap, and where the static structure
of the interface is essentially that of a pure homopolymer interface.
The results are compared to self consistent field theory calculations
for a single copolymer in a homopolymer interface. 
Our paper is organized as follows: The next section describes the simulation 
model and method, and ends with a few comments on the self consistent field 
theory. The results are presented in section three. In particular, we discuss 
segment density profiles, chain and bond orientations, and profiles for the
number of interacting contacts between monomers. We summarize and conclude 
in the last section.

\section{The Bond Fluctuation Model}

With the presently available computational resources, molecular modelling
of the phase behavior in polymer blends in atomistic detail is far
beyond feasibility. Fortunately, many important features of such
systems are already apparent in coarse grained models\cite{B3}, such as the
bond fluctuation model on a cubic lattice\cite{CK}. The latter models polymers
as chains of $N$ effective monomers, which occupy each a cube of
8 neighboring sites and are connected by bond vectors of length
$2, \sqrt{5}, \sqrt{6}, 3$, or $\sqrt{10}$ in units of the lattice spacing 
$a_0$. One such cube represents a group of $n \approx 3-5$ 
chemical monomers. Hence a total chain length of 32, as has been used here, 
corresponds to a degree of polymerization of approx. 100-160 in a real 
polymer. At volume fraction $\phi=0.5 \: a_0^{-3}$ or monomer density 
$\rho=1/16 \: a_0^{-3}$, the model reproduces many important properties of 
dense polymer melts, e.g.  single chain configurations show almost ideal 
Gaussian chain statistics, the single chain structure factor follows a 
Debye function, and the collective scattering function has the 
experimental form\cite{bfm}.

The relative repulsion between monomers of type $A$ and $B$ is modelled
by introducing symmetric energy parameters
$\varepsilon_{AA}=\varepsilon_{BB}=-\varepsilon_{AB}=-k_B T \varepsilon$,
which describe the pairwise interaction between monomers at distances of less 
than $\sqrt{6} \: a_0$. At $\varepsilon = 0.1$,
the interactions between monomers are dominated by the effect of
excluded volume, and the mixture can be described as a weakly perturbed
athermal melt; in particular, the equation of state and the
compressibility are almost not affected by the presence of 
the interactions\cite{MP}.
From extensive previous study of this model, the relation of these model 
parameters to commonly used parameters in polymer theories is well known.
At chain length $N=32$, the statistical segment length $b$ is given
by $b=3.05 \: a_0$\cite{MBO}, the radius of gyration is $R_g = \sqrt{N/6} \: b 
\approx 7 \: a_0$, the compressibility is 
$k_B T \kappa = 3.9 \: a_0^3$\cite{MP}, and the Flory Huggins parameter 
$\chi$ can be calculated using $\chi = 2 z \varepsilon$, where 
$z=2.65$ is the effective coordination number in the bulk,
{\em i.e.}, the average number of {\em inter}chain contacts of a 
monomer\cite{MB}.

The interfacial properties were studied in a $L\times D \times L$ geometry at
system size $D=64$ and $L=512$. The dimensions of the system were chosen such
that the width $D$ of the slab is much larger than the gyration radius $R_g$,
{\em i.e.}, almost ten times as large. The boundary conditions are
periodic in $x$ and $z$ direction and ``antiperiodic'' in the $y$ direction,
{\em i.e.}, $A$-chain parts leaving the right part of the simulation box reenter
it on the left side as $B$-chain parts and vice versa. Since no mechanism
fixes the interface at a certain position, the interface position is subject 
to diffusion due to thermal fluctuations. We choose the coordinate system 
such that the origin of the $y$-axis is at the center $y_0$ of the interfacial 
profile, which we determine via\cite{SB}
\begin{equation}
| \sum_{y_0-20}^{y_0+20} m(y) | = \mbox{min.} 
\end{equation}
Here $m=\rho_A-\rho_B$ is the order parameter of the demixing transition 
and the relative monomer densities are defined by 
$\rho_{A,B} = \phi_{A,B}/\phi$, where $\phi_A$ and $\phi_B$
are the volume fractions taken by $A$ and $B$ monomers. 
The simulation box contains 32768 polymers of equal chain length $32$.
As initial configuration, we choose a relaxed configuration of a pure
homopolymer interface\cite{MBO}, randomly pick 1024 chains with the
center of mass at distances of less than  $\pm \delta$ 
from the interface with $\delta = 3$ or $9$, and turn them into copolymers. 
No effect of the choice of $\delta$ on the results has been found. 
The simulation algorithm
involves random hopping of randomly chosen monomers by one lattice unit
with Metropolis probability, but no grandcanonical moves, {\em i.e.}, the 
number of copolymers remains fixed. 
After an initial equilibration time of 
$2.5\cdot 10^5$ attempted moves per monomer (AMM), the
concentration of copolymers in the bulk at $\varepsilon = 0.1$ is 
$0.05 \% \pm 0.01 \%$
-- estimates from grand canonical bulk simulations suggest
that it should be around $ 0.04 \%$\cite{marcus}. 
We average over 86 configurations in total, where the data for averaging 
are taken  every $10^4$ AMM.
The area covered by one copolymer can be roughly estimated by $\pi R_{g,b}^2$,
where $R_{g,b}^2 = b^2 N/12$ is the gyration radius of one copolymer block. 
Hence $1024$ copolymers cover approximately 
$30 \%$ of the total interface area $512 \times 512$, {\em i.e.},
the system is well in the diluted regime.

We close this section with a brief comment on the self consistent field 
calculations. In a previous paper, we have compared
the properties of homopolymer interfaces with the predictions of
self consistent field theories for completely flexible and semiflexible
chains with different chain rigidities\cite{SM}. The qualitative agreement 
between theory and simulation was over all very good. The main quantitative 
discrepancy was found in the interfacial width -- at chain length $32$, the 
self consistent field theory underestimates the profile widths by at least a 
factor of $2$. This effect could not be explained by capillary waves alone, 
but seemed to be a consequence of the short chain length.
At temperatures sufficiently below the critical point, so that critical 
fluctuations do not affect the interface any more, the interfacial width gets
already comparable to the screening length of the excluded volume, 
on which length scale chains can not be treated as pure random walks.
This leads to wrong predictions of the size of the interface. However,
theory and simulations still agree quantitatively for other quantities,
e.g. the reduction of the total density at the interface.

Our present calculations are based on this work. We consider single wormlike
copolymer chains at an interface of wormlike homopolymers. Chains are 
represented by space curves $\vec{r}(s)$ with $s$ varying from 0 to 1, and 
the single chain partition function of a copolymer is given by
\begin{equation}
{\cal Z} = \int \widehat{\cal D}\{\vec{r}(\cdot)\} \exp \Big[
-\int_0^{1/2} ds W_A(\vec{r}(s)) - \int_{1/2}^1 ds W_B(\vec{r}(s)) \Big],
\end{equation}
where $W_i(\vec{r})$ is the self consistent field acting on a monomer of
type $i$ in a homopolymer interface. Each space curve is assigned
a statistical weight in the functional integral, 
\mbox{$\widehat{\cal D} \{\vec{r}(\cdot)\} = 
{\cal D} \{\vec{r}(\cdot)\} {\cal P}_W \{\vec{r}(\cdot)\}$} with\cite{KP}
\begin{equation}
{\cal P}_W \{ \vec{r}(\cdot) \} = {\cal N}
\prod_s \delta(\vec{u}^2-1)
\exp[ - \frac{\eta}{2 N } \int_0^1 ds | \frac{d\vec{u}}{ds}|^2 ],
\end{equation}
where $a$ is the fixed monomer length, $\eta$ a
dimensionless stiffness parameter, $\vec{u} = \frac{d\vec{r}}{ds} /(N a)$ 
the dimensionless tangent vector constrained to unity by the delta function,
and ${\cal N}$ the normalization factor. Here we choose $\eta=0.5$, 
the value which best reproduces bond orientations in pure homopolymer 
interfaces in the bond fluctuation model\cite{SM}. The self consistent 
fields $W_i$ for such an interface of semiflexible homopolymers have 
previously been determined numerically in Ref. \cite{SM},
based on the Helfand type free energy functional
\begin{equation}
\beta {\cal F} = \rho \int d\vec{r} \{
\chi \Phi_A \Phi_B + \frac{1}{2 \rho k_B T \kappa} 
(\Phi_A + \Phi_B - 1)^2 \}
\end{equation}
with the total bulk monomer density $\rho$, the relative monomer 
densities $\Phi_i(\vec{r}) = \rho_i(\vec{r})/\rho$ ($i=A$ or $B$),
the Flory Huggins parameter $\chi$, and the compressibility $\kappa$.

The distribution of copolymer segments is calculated by solving appropriate
diffusion equations for the end segment distributions
$Q(\vec{r},\vec{u};s)$ and $Q^+(\vec{r},\vec{u};s)$ 
for chain parts of length $sN<N$, which begin on the $A$ side ($Q$) or 
the $B$ side ($Q^+$) of the copolymer\cite{MF,MM}.
From those one can calculate the density of monomers at position $s$ with 
orientation $\vec{u}$ via $\Phi(\vec{r},s) = Q(\vec{r},s) Q^+(\vec{r},1-s)$ or
$\Phi(\vec{r},\vec{u}) = Q(\vec{r},\vec{u},s) Q^+(\vec{r},-\vec{u},1-s)$.
The numerical treatment is facilitated by expanding the functions 
$Q$, $Q^+$ in Legendre polynomials and including only the three lowest 
moments\cite{SM}.

\section{Results}

All results presented here were obtained at $\varepsilon = 0.1$ or 
$\chi N = 17$, {\em i.e.}, the system is far from the critical point 
($(\chi N)_c = 2.4$~\cite{bfm,MB}) in 
the strong segregation regime. Simulations were also performed
at $\varepsilon = 0.05$; the results are qualitatively the same,
but the effects are less marked. 

In the following,
lengths are given in units of $w_{SSL}=b/\sqrt{6 \chi}$, the
interfacial width of a homopolymer interface in the mean field strong 
segregation limit. The radius of gyration in these units is 
$R_g = 4.2 \: w_{SSL}$, and the slab thickness 
$D =  37.4 \: w_{SSL}$.  

\subsection{Density profiles and segment distributions}

Figure 1 shows profiles of $A$ and $B$ monomer densities in systems with and 
without copolymers, profiles of just the $A$ and $B$ blocks
of the copolymers, and total density profiles. 
As expected in the diluted
regime, the $A$ and $B$ monomer density profiles are hardly affected
by the presence of the copolymers.
The interface as a whole is still very similar to a pure homopolymer interface. 
Due to the finite compressibility of the blend, the total density is slightly 
reduced in the interfacial region\cite{MBO}. The distributions of copolymer 
monomers $A$ and $B$ agree qualitatively with the experimental results and 
confirm the simple picture presented in the introduction: Monomers of type $A$ 
are more concentrated in the $A$-rich phase, Monomers of type $B$ in the $B$ 
rich-phase. The distribution is rather broad, a fair portion of the $A$ 
monomers sticks into the $B$-rich phase and vice versa.

Distributions of single chain segments are shown in Figure 2 and 3. For
homopolymers, one finds a relative enrichment of chain ends $\rho_e$ at the
interface, whereas the concentration $\rho_{1/2}$ of middle segments (the 16th 
and 17th monomer) is comparatively low there. The total density profile 
$\rho_h$ is best reproduced by the density profiles $\rho_{1/4}$ of the 
segments at one and three fourth of the chain (the 8th and 25th monomer). 
Looking at the higher concentration of chain ends at the interface, one is 
lead to suspect that homopolymers tend to form loops with two ends at the 
interface. However, this is not the case, as can be seen from the distribution 
$\rho_{e-e}$ of midpoints between the two ends of homopolymers. 
It is strongly reduced at the center of the interface and enhanced at the
distance of one gyration radius from there (Figure 2).

Copolymer segment densities show the inverse trend: The middle segments
concentrate at the interface, whereas the chain ends stretch out into 
their favorite bulk phase. Hence our results are in qualitative agreement 
with the experimental findings of Russell et al~\cite{SMR}. As in the case of
homopolymers, the total density profiles of $A$ and $B$ monomers are almost 
identical with the distribution of the 8th and 25th monomer, respectively,
{\em i.e.}, the distribution of segments in the middle of the $A$ or $B$ block.
Note that the peak of the distribution of middle segments $\rho_{1/2}$ is
relatively broad, broader than the radius of gyration, hence
they are not strongly confined to the interface as
assumed by the Leibler theory (Figure 3).
The results of the self consistent field theory are shown in the
inset. The fact that the SCFT underestimates the interfacial width also
leads to quantitative discrepancies in the distribution of
copolymer segments. However, the qualitative agreement is very good,
and in the wings of the profile one even reaches quantitative agreement.

\subsection{Chain and bond orientations}

Next we discuss the orientational properties of the polymers. It is
instructive to consider separately the orientations of single bonds, of
chain segments, and of whole chains. Chains with rather weakly oriented
single bonds can still be strongly oriented as a whole, as found by
M\"uller et al for homopolymers at a homopolymer interface\cite{MBO,SM}. 

The orientation of whole chains involves two different factors. First, 
chains may be oriented without volume changes, {\em i.e.}, the total gyration
radius or end-to-end vector remains unaffected by the 
orientation. A weak orienting field is sufficient to bring 
about orientation of this kind. Second, chains may get compressed or
stretched in one direction. As Figure 4 illustrates, the latter effect
dominates close to an interface. The mean squared components of the 
end-to-end vector in directions parallel ($x$,$z$) and perpendicular ($y$) 
to the interface are shown for homopolymers and copolymers. The 
components parallel to the interface hardly vary throughout the system.
Perpendicular to the interface, the end-to-end vector of homopolymers is 
reduced in the interfacial region, as is already the case in pure homopolymer 
systems. Homopolymers are thus squeezed perpendicular to the 
interface, and get effectively oriented parallel to the interface. 

Copolymers show the inverse behavior, they stretch in the direction 
perpendicular to the interface. As obtained both from self consistent
field calculations and from the simulations, the effect is very strong 
for copolymers centered at about one to two radii of gyration 
away from the interface, and much weaker for 
those located in the wings of profile or
at the middle of the interface (Figure 4). 
One can picture the latter as
consisting of two almost independent homopolymer blocks, which hardly feel the
effect of being linked together at one end. 
Indeed, the vector connecting the 
ends of the single $A$ or $B$ blocks is on average hardly oriented 
(Table 1). 
Only the blocks
centered deep in their majority phase ($y/w_{SSL} \approx 7$)
stretch perpendicular to the
interface, since they are pulled towards the interface by the other
copolymer end (Figure 5).
Note that Figure 5 also demonstrates, how minority blocks ($A$ blocks in
the $B$ phase, $B$ blocks in the $A$ phase) slightly shrink by a factor of 
approximately 0.8 due to the hostile environment.
The vector $\vec{D}_{AB}$ connecting 
the centers of mass of the $A$ and $B$ blocks is on average oriented and 
strongly stretched in the negative $y$ direction (Table 1).

In sum, single blocks are mostly not oriented at
all; the perpendicular orientation 
of whole copolymers results from the arrangement of the two constituent 
blocks. Diblock copolymers can be pictured as dumbbells\cite{WV} consisting of 
two mildly perturbed homopolymer coils. Similar copolymer shapes have already
been found by Binder and Fried in simulations of block copolymers in
the disordered phase, far above the ordering transition\cite{BF}.

Consequently, one would expect that the only region, where the local 
conformation of a diblock differs significantly from a homopolymer 
conformation, is the region close to the link which connects the two blocks.
We study this in more detail by looking at the orientation of single 
bonds $\vec{b}$. In order to do so, it is useful to define a bond 
orientation parameter 
\begin{equation}
q(\vec{r}) =
\frac{ \langle b_y^2 \rangle - \frac{1}{2}
(\langle b_x^2 \rangle + \langle b_z^2 \rangle) }
{\langle \vec{b}^2 \rangle}.
\end{equation}
Negative $q$ implies orientation parallel to the interface, 
positive $q$ perpendicular orientation. 

The orientations of homopolymer bonds are shown in Figure 6. Like whole 
chains, but to a much lesser extent, bonds tend to align themselves parallel 
to the interface. Unlike the end-to-end vector of whole chains, the 
average squared bond length $\langle \vec{b}^2 \rangle$ varies by less than 
$0.15 \%$ 
throughout the system (not shown here), hence the bond length is quasi
fixed, the bonds get oriented without compression. The effect is 
strongest in the middle of the chains and qualitatively the same, but weaker 
for end bonds.
The bond orientation parameter can be compared to the average orientation
of the tangent vector $\vec{u}$ predicted by the self consistent field theory
$q \widehat{=} \frac{1}{2}(3 \langle u_y^2 \rangle - 1)$. The results
show the same trend as the simulation data for the middle bonds, but do not
predict any effect for the end bonds (Figure 6, inset), unlike what is seen
in the MC simulations. This reflects the difference between the 
tangent vector at the end of a continuous space curve, as assumed by SCFT, 
and an end bond connecting two discrete statistical monomers of a relatively 
short chain, as it is the case in our MC simulation.

In contrast to homopolymer chains, MC data reveal a rather strong dependence
of the orientation of bonds in copolymers on their position within the chain (Figure 7a). 
The bond which links
the two blocks is preferably oriented perpendicular to the interface -- 
the stronger, the further away its location from the center of the interface.
Already the bonds next to the link bond show a much smaller effect.
The larger the distance along the chain from the link bond, the stronger
the tendency of parallel alignment at the center of the interface
becomes, and the further it reaches out into the wings of the profile. 
Hence bonds in the middle and at the end of an $A$ or $B$ block behave very much 
like homopolymer bonds.
We note {\em en passant} that, compared to the other bonds, 
the link bond is stretched by 
$\sim 4 \%$ 
due to the relative repulsion 
between the adjacent $A$ and $B$ monomers. 

The results obtained by SCFT
show the same trends as the simulation data (Figure 7b), and agree
qualitatively, but the same remarks as above apply. Thus, SCFT does
not reproduce the slight tendency of parallel alignment which is
found in the MC data at the ends of the blocks.
The orientational parameter of the link bond at the center of the
profile is $q \approx +0.025$ according to both MC results and SCFT.
However, SCFT underestimates the average orientation
(averaged over link bonds in the region shown in Figure 7a) by $\sim 30 \%$. 
At distances of several gyration radii $R_g$ from the
interface, the bond orientation parameter $q$ drops back to zero, 
indicating that copolymers that deep in the bulk are oblivious to
the effect of the interface.
Note that two different length scales are reflected in the profiles of $q$. One
of them, the width of the homopolymer interface, fixes the
width of the central dip in the profiles, the other one, the gyration radius,
determines the overall width of the region with nonzero $q$.

\subsection{Self and Mutual Contacts}

Another instructive quantity in the bond fluctuation model is the number
of monomer ``contacts'', {\em i.e.}, the number of interacting monomer pairs.
It is expedient to distinguish between ``self contacts'', {\em i.e.}, contacts 
between monomers belonging to the same chain (``intrachain contacts''), and 
``interchain contacts'' of monomers from 
two 
different chains. The self contacts 
provide additional information on the conformation of single chains, and the 
interchain contacts on the arrangement of chains relative to each other.

Figure 8 compares the number of self contacts per monomer $N_{i,self}/\rho_i$ 
in homopolymer ($i = h$) and copolymer ($i = c$) chains. 
The main contribution to $N_{i,self}/\rho_i$ 
comes from the two direct neighbors of a monomer in the chain, thus 
$N_{i,self}/\rho_i$ is mostly slightly larger than 2. 
Note however that direct neighbors do not always interact with each other,
since the maximum bond length $\sqrt{10} \: a_0$ is larger than the
range of the interactions $\sqrt{6} \: a_0$.
The number of self contacts in homopolymers is almost constant throughout
the system. It is slightly enhanced in the wings of the profile and 
decreases again at the center, where $A$ and $B$ meet -- suggesting
that homopolymer chains tend to loop away from the interface, such that
there is no room for intrachain contacts at the center, but that the number
of contacts increases right next to it. In copolymers, one finds the
opposite behavior -- the relative number of intrachain contacts is higher at
the center of the interface than in homopolymers, but diminishes
rapidly as one moves away from the interface. This trend is promoted by
three factors. First, far from the interface copolymers stretch towards
the interface, and the number of self contacts associated with backfolding
goes down. Second, the stretching goes along with a higher occupation of
the $(0,3,0)$ bond, which is outside of the interaction region,
hence monomers gradually lose contact to their 
direct neighbors. Third, the further away one moves from the interface, 
the greater becomes the contribution of chain end monomers, which only have 
one direct neighbor in the chain. Obviously, it would be highly desirable 
to study separately the contributions to $N_{i,self}/\rho_i$ from monomers which are 
direct neighbors in a chain, and from monomers which are further apart from 
each other. Unfortunately, such a distinction was not possible due to limited
memory space.

Whereas the number of self contacts reflects the conformational properties
of a chain, the demixing of $A$ and $B$ chains in the melt is essentially
driven by the interchain contacts. The number of interchain contacts 
per monomer can be interpreted as an effective coordination number $z_{eff}$.
In homogeneous systems, the identification $\chi = 2 z_{eff} \varepsilon$
makes contact with the Flory-Huggins theory\cite{MB}.
In inhomogeneous systems, the effective coordination number is position
dependent. As long as the finite range of interactions is neglected, mean 
field theory simply asserts that it is proportional to the local density 
of monomers: 
\begin{equation}
\label{z}
z_{i,eff}(\vec{r}) = \frac{N_{i,inter}(\vec{r})}{\rho_i(\vec{r})} 
\propto \rho(\vec{r})
\end{equation}
A more elaborated mean field approach predicts
for a system which is inhomogeneous in one direction $y$
\cite{F2},
\begin{equation}
z_{eff} \propto \rho(y) + \frac{1}{2} k^2 \frac{d^2}{dy^2} \rho(y) \quad
\mbox{with} \quad 
k^2 = \frac{\int d\vec{r} \gamma(\vec{r}) V(\vec{r}) y^2}
           {\int d\vec{r} \gamma(\vec{r}) V(\vec{r})},
\end{equation}
where $V(\vec{r})$ is the integrable part of the interaction potential
({\em i.e.}, excluding the hard core part) and $\gamma(\vec{r})$ the normalized
pair correlation function. Since the interaction region extends no further 
than $|y| \le 2 a_0$, an upper bound for the factor $k^2$ is given by
$k^2 \le 4 a_0^2$. Figure 1 shows that $\rho''(y)/\rho(y) \le 0.01/a_0^2$. 
Hence the expected deviation from the simple proportionality law (\ref{z}) 
is of order $\le 2 \%$. Furthermore, one would expect that the effective 
coordination number does not depend on whether a monomer belongs to a 
copolymer or a homopolymer.

In contrast to these considerations, the Monte Carlo data reveal a much
lower dip of the reduced coordination number $z_{eff}(y)/\rho(y)$ 
(Figure 9) at the center of the interface. The discrepancy between 
mean field assumption and simulation data becomes even more manifest when 
looking at the number of contacts $N_{AB}$ between $A$ and $B$ monomers. 
From the usual mean field assumption, one would expect 
$N_{AB} \propto \rho_A \cdot \rho_B$ (Figure 10). The simulation data
however show that $N_{AB}/(\rho_A \rho_B)$ is increased by a factor of
$1.5$, compared to the bulk value, at the distance of approximately two 
radii of gyration from the interface, where the chains are 
stretched (cf. Figure 5). 
At the center of the interface, on the other hand, it is decreased
by $ 25 \% $.

One can deduce
that chains are compactified in the interfacial region. Interestingly, 
this holds particularly for copolymer chains. We recall that the number 
of {\em intra}chain contacts in copolymers is enhanced at the interface, 
whereas the {\em inter}chain contacts are obviously suppressed. Thus
copolymers offer relatively little ``active surface'' for interaction with
external monomers in the interfacial region. 
In contrast, the relative number of external contacts increases further away
from the interface. Hence the active surface per monomer will 
presumably increase with the copolymer chain length. 
This may be an additional reason why short copolymers are relatively poor
amphiphiles.

\subsection{Summary}

We have studied the ternary system of $A$ and $B$ homopolymers and
symmetric $AB$ diblock copolymers close to an $A$/$B$ interface in
the dilute regime, where copolymer coils do almost not overlap with each other,
by Monte Carlo simulations of the bond fluctuation model.
In this regime, the structural properties of the interface are
not altered with respect to the pure homopolymer interface -- in
particular, we find no significant broadening of the interfacial
width, as is expected at higher copolymer concentrations.
According to our simulation results, copolymers resemble oriented 
dumbells with the $A$ block sticking into the $A$ rich phase, and 
the $B$ block sticking into the $B$ rich phase. The conformations
of single blocks are not very different from conformations
of pure homopolymer coils. Single blocks tend to
orient themselves parallel to the interface, like homopolymers, whereas
copolymers as a whole are oriented perpendicular to the interface.
When looking at the orientations of single bonds, we find that the
orientational properties of the central bond, which links the two blocks, 
differ distinctly from those of a homopolymer bond, but that already
the neighboring bonds behave very similar to homopolymer bonds. Hence the 
chain loses the memory of the link very rapidly. Our results are reproduced 
qualitatively, but not quantitatively, by a self consistent field 
calculation.  Furthermore, we found that copolymers are unusually
compact in the interfacial region, {\em i.e.}, monomers have more contacts to 
monomers of the same chain, and fewer contacts to monomers of different 
chains than monomers in the bulk phase. As a result, the 
copolymer-copolymer interchain contacts are particularly suppressed at the 
interface. This should become important as the concentration of copolymers
is increased. Future work will be concerned with the structure of
interfaces with higher content of copolymers and with the transition
between different regimes, from the dilute to the wet brush to the dry brush 
regime, until the two phase region breaks down and a lamellar phase emerges.

\section*{Acknowledgements}

We thank H. Weber for invaluable help in solving some computational
problems. K.B. thanks W.L. Mattice for stimulating discussions.
A.W. received partial support from the Deutsche Forschungsgemeinschaft 
under grant No. Bi 314/3-4. Further support from grant No. Bi 314/12-1,
and generous access to the Cray T3D at EPFL is acknowledged.

\newpage
\section*{Table}
\pagestyle{empty}
\begin{table}[h]
\caption{
Orientational properties of copolymer chains. Averages are taken over all 
1024 copolymers and over 86 independent configurations. 
Compared are $x$, $z$ components (together) with $y$ components. 
Lengths are given in units of $w_{SSL}$.
}
\begin{center}
\begin{tabular}{lllcc}
\hline
&& & $i = x,z$& $i = y$	\\ 
\hline
end-to-end vector:& $A$ and $B$ blocks ($N = 16$)& $\langle R_i^2 \rangle$ 
& 15.3 		& 15.6		\\
& whole copolymer chain ($N = 32$)& $\langle R_i^2 \rangle$	& 32.7		& 	47.4	\\
\multicolumn{2}{l}{vector from center of block $A$ to center of block $B$}
& $\langle D_{i,AB}^2 \rangle$	& 11.6		& 19.4		\\
& & $\langle D_{i,AB} \rangle$	& 0.0		& -3.8		\\
\hline
\end{tabular}
\end{center}
\end{table}

\clearpage
\newpage

\section*{Figure Captions}

\begin{description}

\item[Figure 1:] 
Monomer density profiles as a function of the distance from the center of the 
interface $y$, in units of $w_{SSL}=b/\sqrt{6 \chi} = 1.71 \: a_0$. 
Profiles are shown for the density
of all monomers ($\rho$), of $A$ and $B$ monomers separately ($\rho_A$ and
$\rho_B$), of just homopolymer monomers ($\rho_h$), and of $A$ and $B$ 
monomers belonging to a copolymer block ($\rho_{A,c}$ and $\rho_{B,c}$). 
Also shown for comparison are $A$ and $B$ monomer profiles in a pure
homopolymer system ($\rho_A^o$ and $\rho_B^0$, taken from \cite{MBO}).

\item[Figure 2:]
Homopolymer segment density profiles vs. $y/w_{SSL}$ in units of 
$w_{SSL}=b/\sqrt{6 \chi}$. Profiles are shown for the density
of monomers in the middle of the chain ($\rho_{1/2}$), at the
end of the chain ($\rho_e$), at one and three fourth of the chain
($\rho_{1/4}$), and of all homopolymer monomers ($\rho_h$). Also
shown is the distribution of midpoints between the two ends of homopolymers,
$\rho_{e-e}$.
Units of densities are the total bulk concentration of monomers
$\rho_b$, or $\rho_{sb} = \rho_b/16$ (as indicated).

\item[Figure 3:]
Copolymer segment density profiles vs. $y/w_{SSL}$ in units of 
$w_{SSL}=b/\sqrt{6 \chi} = 1.71 \: a_0$. Profiles are shown for the density
of $A$ and $B$ monomers in the middle of the chain ($\rho_{1/2}$, squares), at the
end of the chain ($\rho_e$, circles), at one and three fourth of the chain
($\rho_{1/4}$,diamonds), and of all copolymer monomers ($\rho_c$, broken line). 
Inset shows results of the self consistent field theory for $A$ monomers. Full lines
show the predictions for segment density profiles $\rho_{1/2}, \rho_e,
\rho_{1/4}$, broken line shows total density profile $\rho_c$, and symbols 
compare with MC results (symbols like above).
Units of densities are $\rho_b / 2$, or $\rho_{sb}=\rho_b/32$ (as indicated).

\item[Figure 4:]
Mean square end-to-end vector components $\langle R_i^2 \rangle$ with $i=x,y,z$
in units of the average bulk value $b^2 N/3$, plotted vs. the distance of 
the center of the end-to-end vector from the interface $y$
in units of $w_{SSL}=b/\sqrt{6 \chi}$. Results are shown for
homopolymers and copolymers, and compared to the prediction 
for copolymers of the self consistent field theory.

\item[Figure 5:]
Mean square end-to-end vector components $\langle R_i^2 \rangle$ ($i = x,z$, 
or $y$)
of the copolymer blocks in their minority phase
($A$ block in $B$ phase, $B$ block in $A$ phase), 
and majority phase ($A$ block in $A$ phase, $B$ in $B$ phase),
in units of the average bulk value $b^2 N/6$, plotted vs. the distance of
the center of the end-to-end vector from the 
interface $y$ in units of $w_{SSL}=b/\sqrt{6 \chi}$.

\item[Figure 6:]
Orientational order parameter $q$ for end bonds (dashed line) and middle bonds 
(the bonds connecting the 16th and 17th monomer, solid line) in homopolymers, 
vs. $y/w_{SSL}$ in units of $w_{SSL}=b/\sqrt{6 \chi}$. Inset shows the
prediction of the self consistent field theory.

\item[Figure 7:]
(a) Orientational order parameter $q$ for end bonds, link bonds (linking 16th 
and 17th monomer), bonds next to link bonds (15th to 16th monomer and
17th to 18th monomer), and bonds in the middle of a block
(8th to 9th monomer and 24th to 25th monomer) in copolymers, vs.
$y/w_{SSL}$ in units of $w_{SSL}=b/\sqrt{6 \chi}$. (b) Prediction
for $q$ of the self consistent field theory for $s = (0,1)$ (end bonds),
$s = 0.5$ (link bonds), $s = (0.25,0.75)$ (block middle bonds).

\item[Figure 8:]
Number of intrachain contacts per monomer $N_{i,self}/\rho_i (y)$ (i = h,c)
vs. $y/w_{SSL}$ in units of $w_{SSL}=b/\sqrt{6 \chi}$ 
for homopolymer and copolymer chains.

\item[Figure 9:]
Normalized effective coordination number $z_{eff}(y)/\rho(y)$
vs. $y/w_{SSL}$ in units of $w_{SSL}=b/\sqrt{6 \chi}$,
for homopolymer and copolymer chains.

\item[Figure 10:]
Normalized number of $AB$ contacts $N_{AB}/(\rho_A(y) \rho_B(y))$, and
number of $AB$ contacts per monomer $N_{AB}/\rho(y)$ (inset), 
vs. $y/w_{SSL}$ in units of $w_{SSL}=b/\sqrt{6 \chi}$.

\end{description}

\newpage

\pagestyle{empty}

\LARGE
\unitlength=1mm
\begin{picture}(150,150)
\put(-20,0){
\psfig{figure=\dir/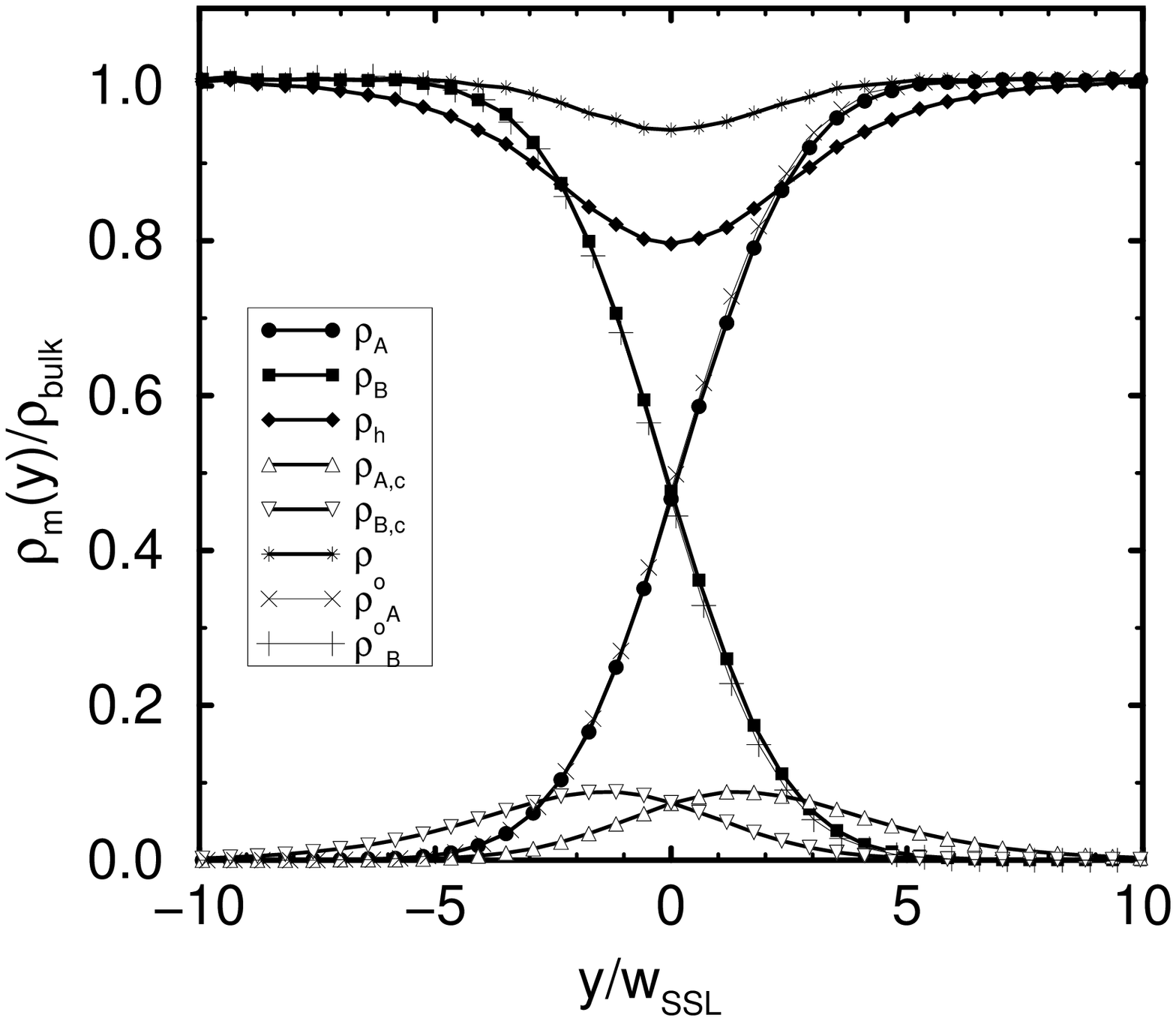,width=160mm,height=140mm}
}
\end{picture}
\vfill
\normalsize
{\tt
\noindent
Figure 1\\
Werner et al
}

\newpage
\pagestyle{empty}

\LARGE
\unitlength=1mm
\begin{picture}(150,150)
\put(-20,0){
\psfig{figure=\dir/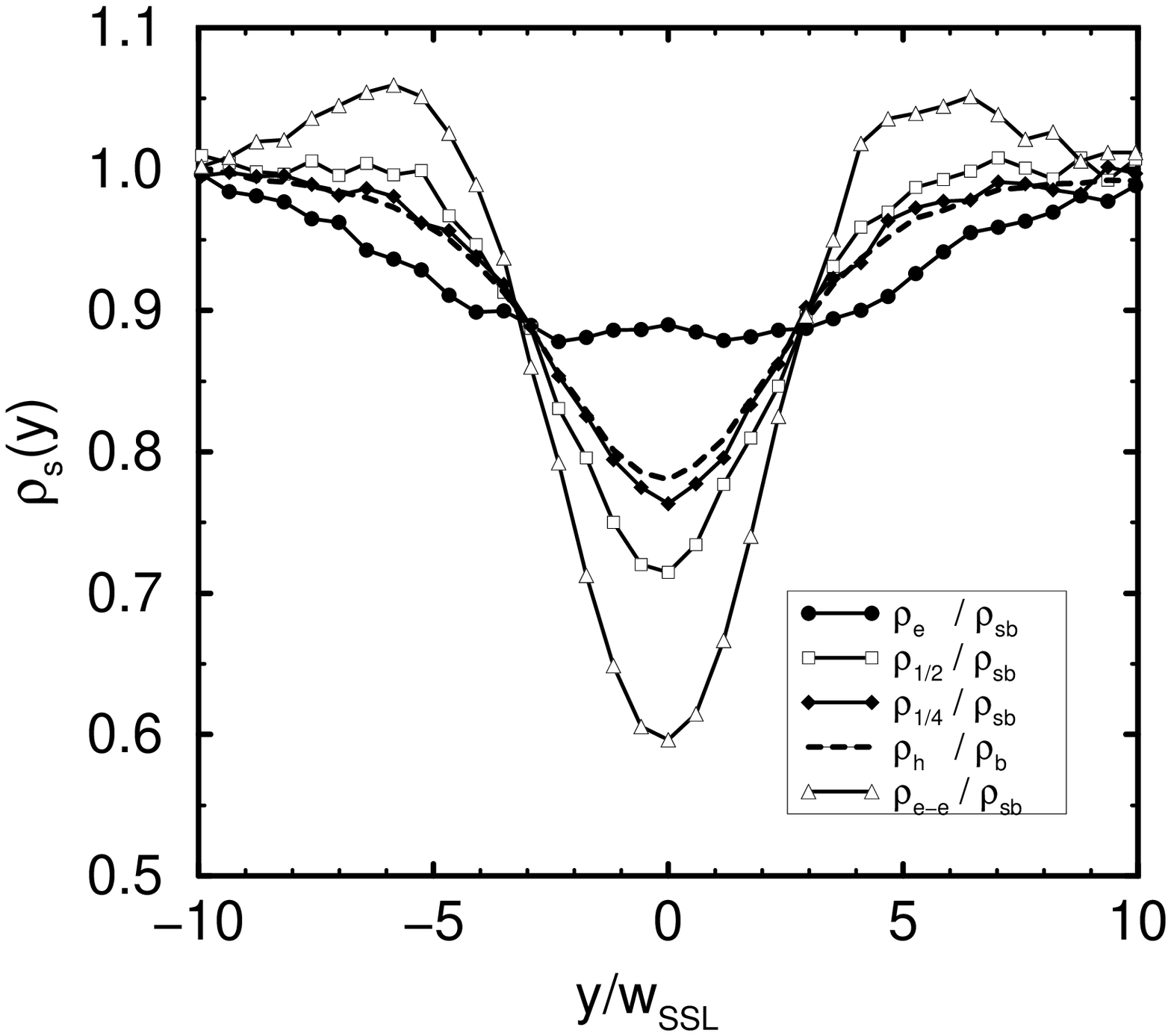,width=160mm,height=140mm}
}
\end{picture}
\vfill
\normalsize
{\tt
\noindent
Figure 2\\
Werner et al
}

\newpage
\pagestyle{empty}

\LARGE
\unitlength=1mm
\begin{picture}(150,150)
\put(-20,0){
\psfig{figure=\dir/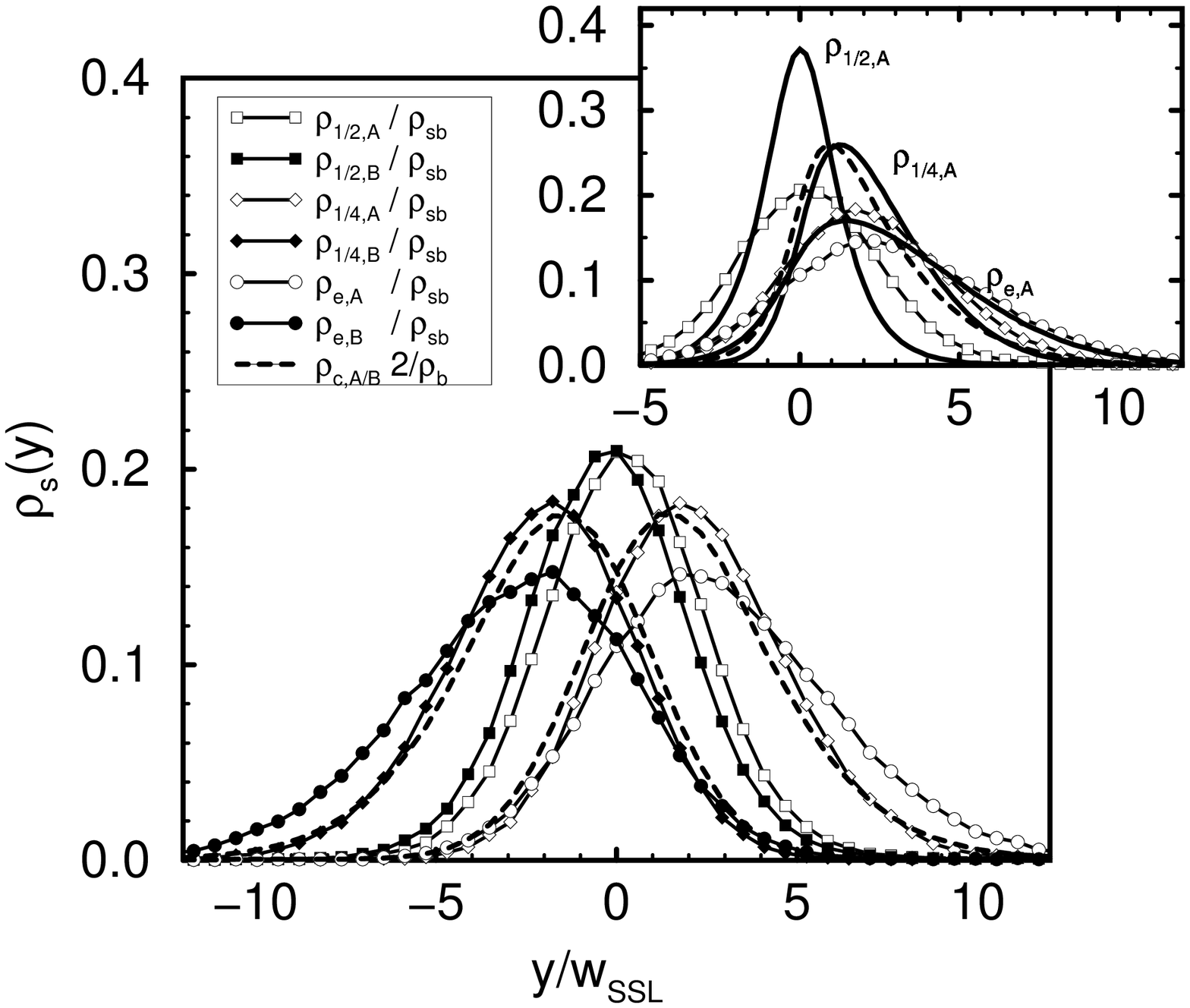,width=160mm,height=140mm}
}
\end{picture}
\vfill
\normalsize
{\tt
\noindent
Figure 3\\
Werner et al
}

\newpage
\pagestyle{empty}

\LARGE
\unitlength=1mm
\begin{picture}(150,150)
\put(-20,0){
\psfig{figure=\dir/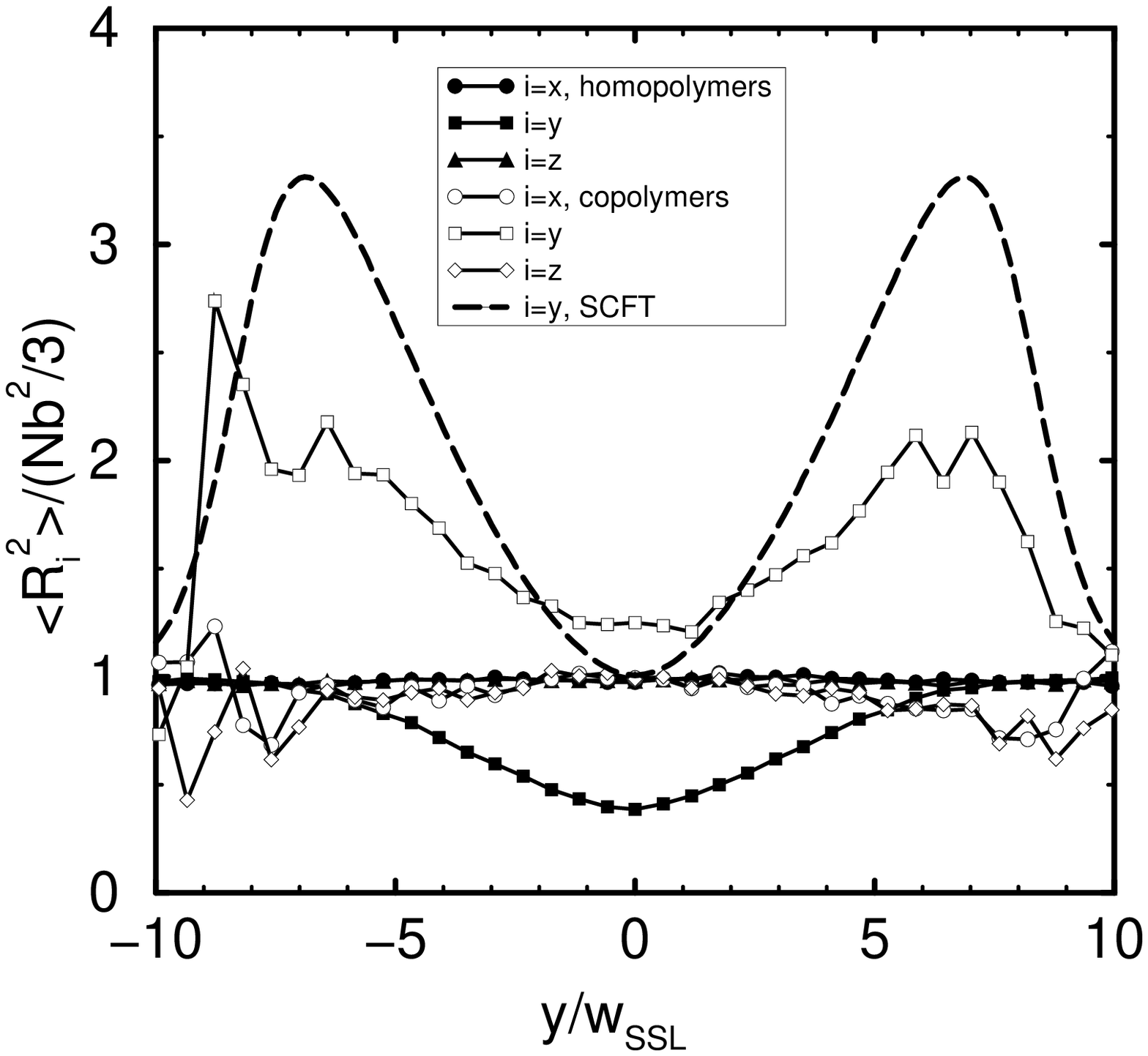,width=160mm,height=140mm}
}
\end{picture}
\vfill
\normalsize
{\tt
\noindent
Figure 4\\
Werner et al
}

\newpage
\pagestyle{empty}

\LARGE
\unitlength=1mm
\begin{picture}(150,150)
\put(-20,0){
\psfig{figure=\dir/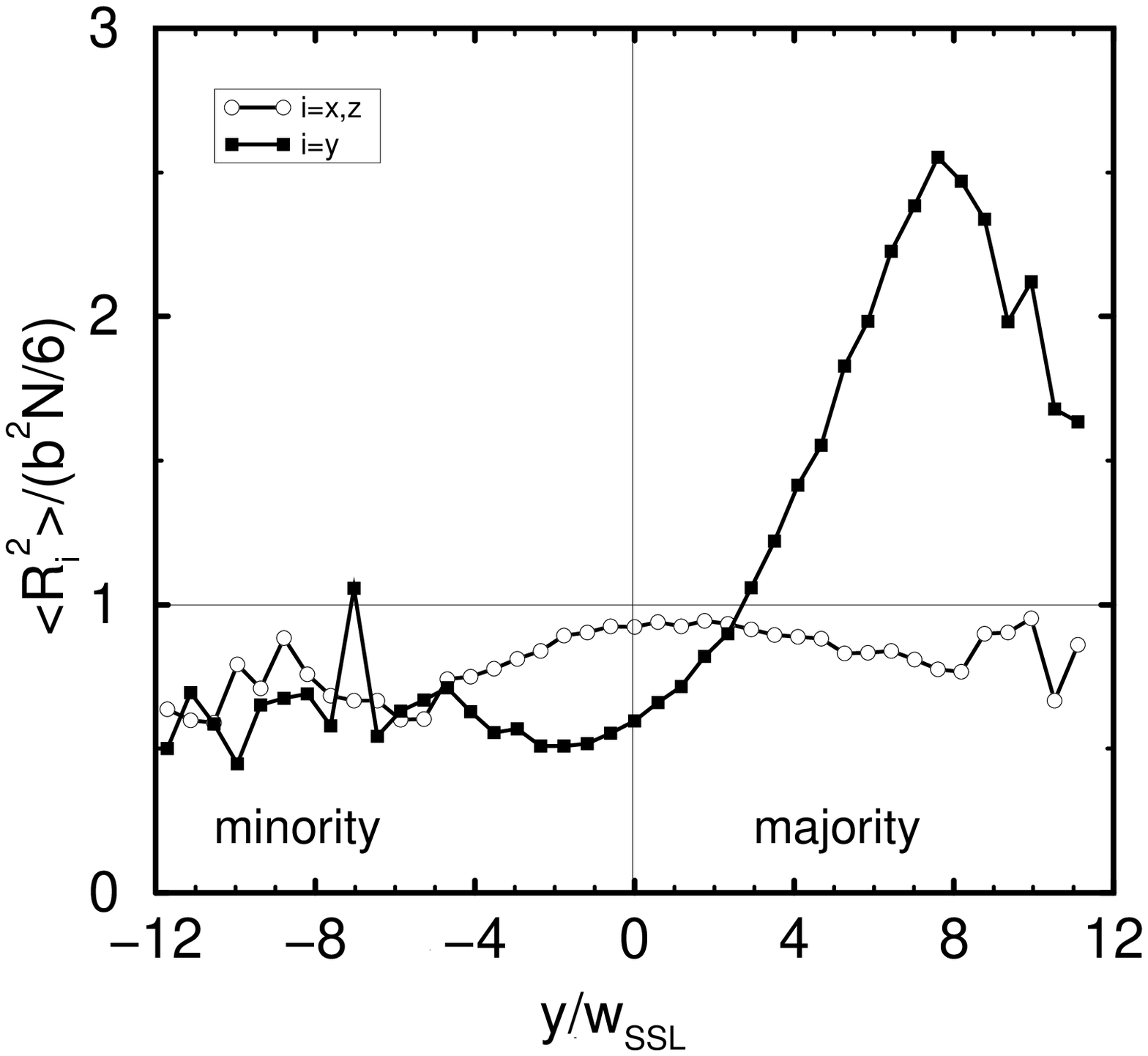,width=160mm,height=140mm}
}
\end{picture}
\vfill
\normalsize
{\tt
\noindent
Figure 5\\
Werner et al
}

\newpage
\pagestyle{empty}

\LARGE
\unitlength=1mm
\begin{picture}(150,150)
\put(-20,0){
\psfig{figure=\dir/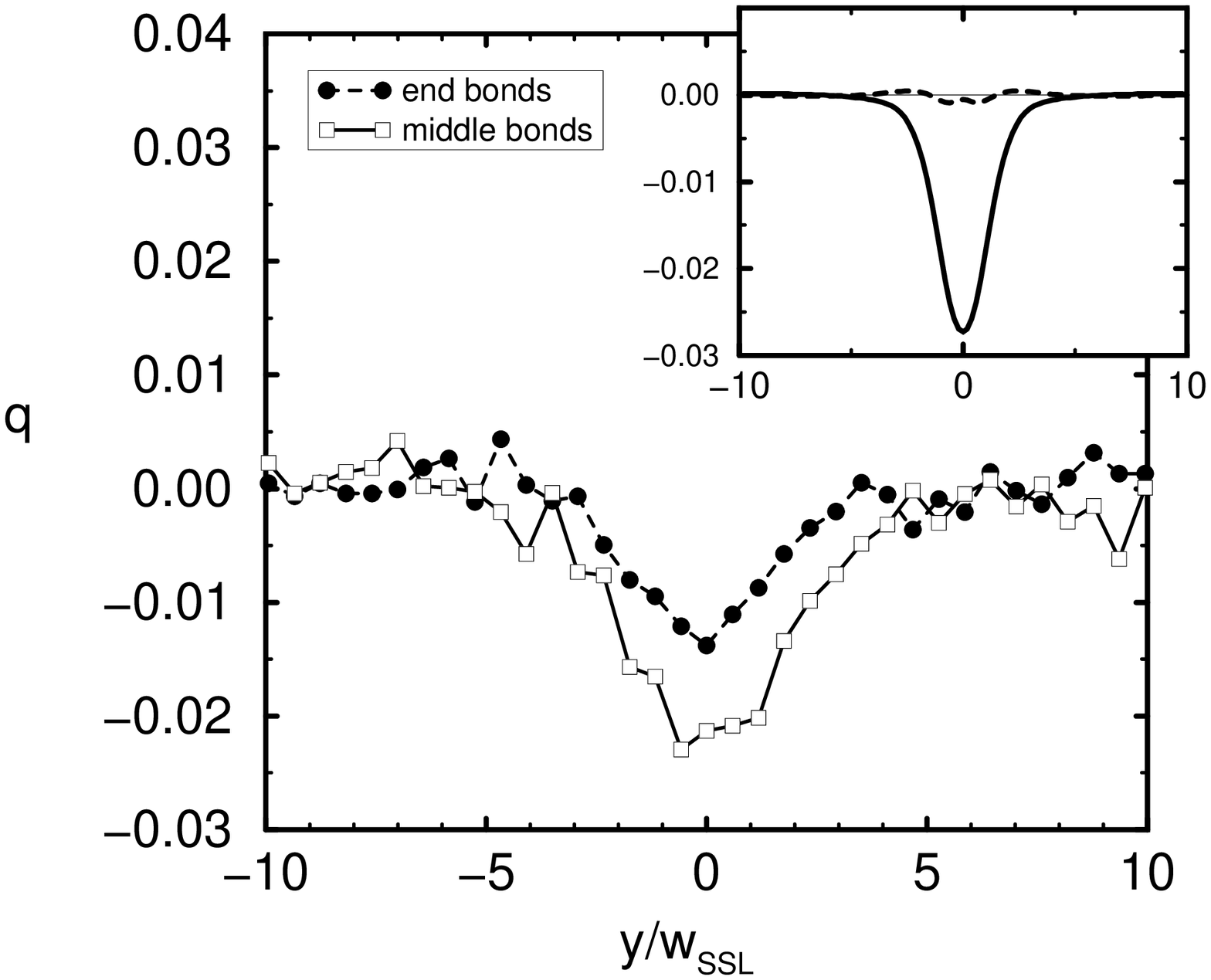,width=160mm,height=140mm}
}
\end{picture}
\vfill
\normalsize
{\tt
\noindent
Figure 6\\
Werner et al
}

\newpage
\pagestyle{empty}

\LARGE
\unitlength=1mm
\begin{picture}(150,150)
\put(-20,0){
\psfig{figure=\dir/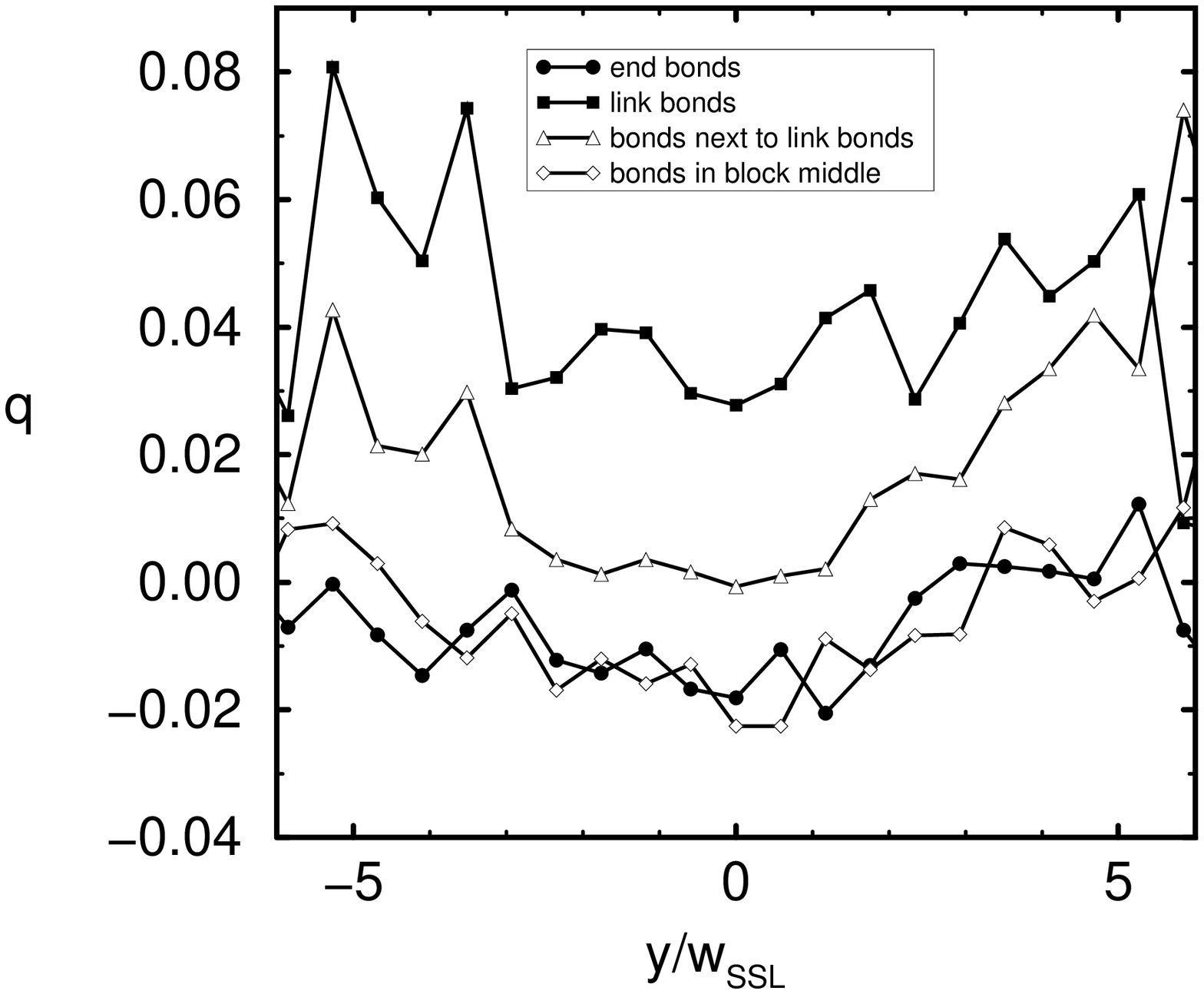,width=160mm,height=140mm}
}
\end{picture}
\vfill
\normalsize
{\tt
\noindent
Figure 7a\\
Werner et al
}

\newpage
\pagestyle{empty}

\LARGE
\unitlength=1mm
\begin{picture}(150,150)
\put(-20,0){
\psfig{figure=\dir/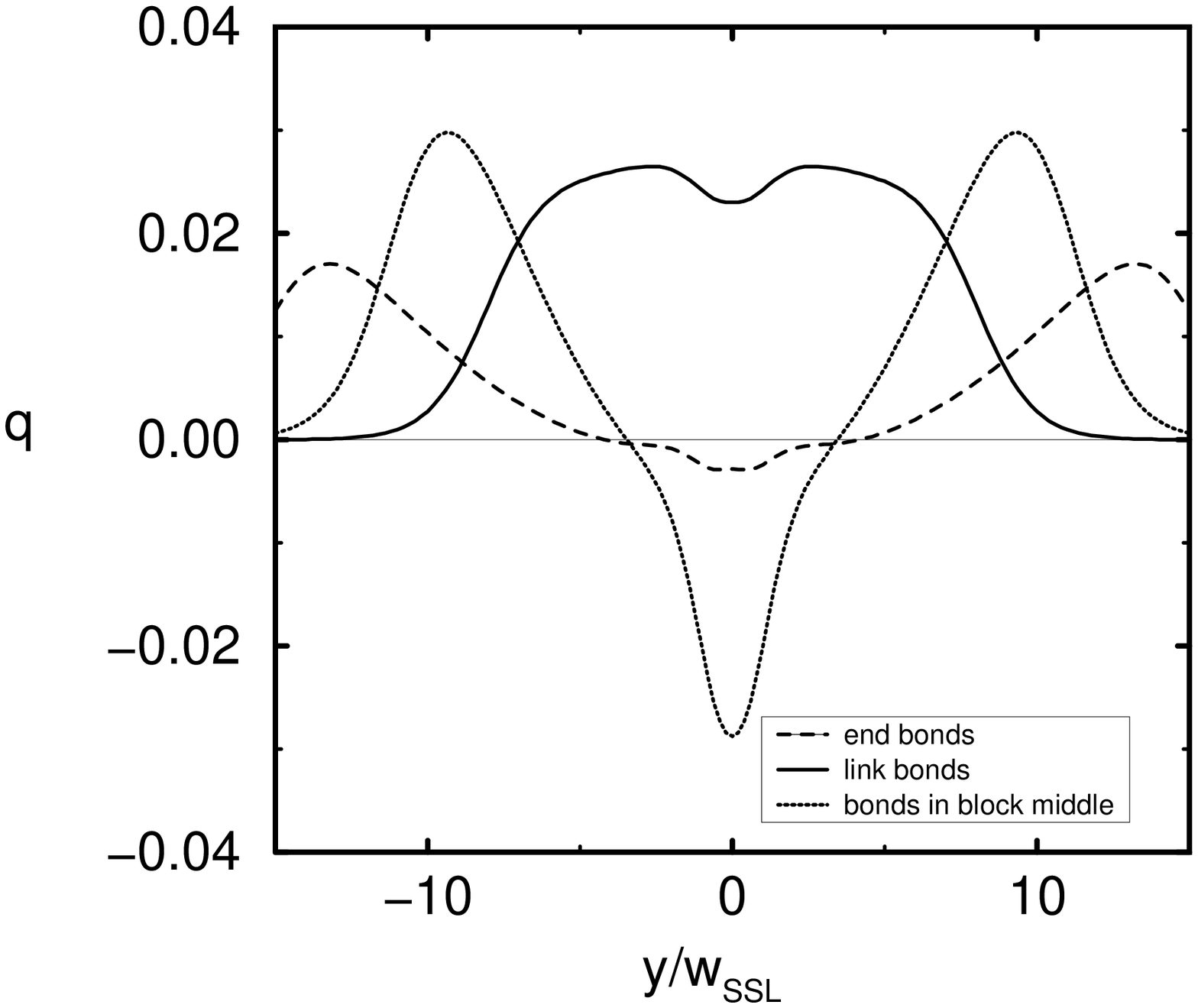,width=160mm,height=140mm}
}
\end{picture}
\vfill
\normalsize
{\tt
\noindent
Figure 7b\\
Werner et al
}

\newpage
\pagestyle{empty}

\LARGE
\unitlength=1mm
\begin{picture}(150,150)
\put(-20,0){
\psfig{figure=\dir/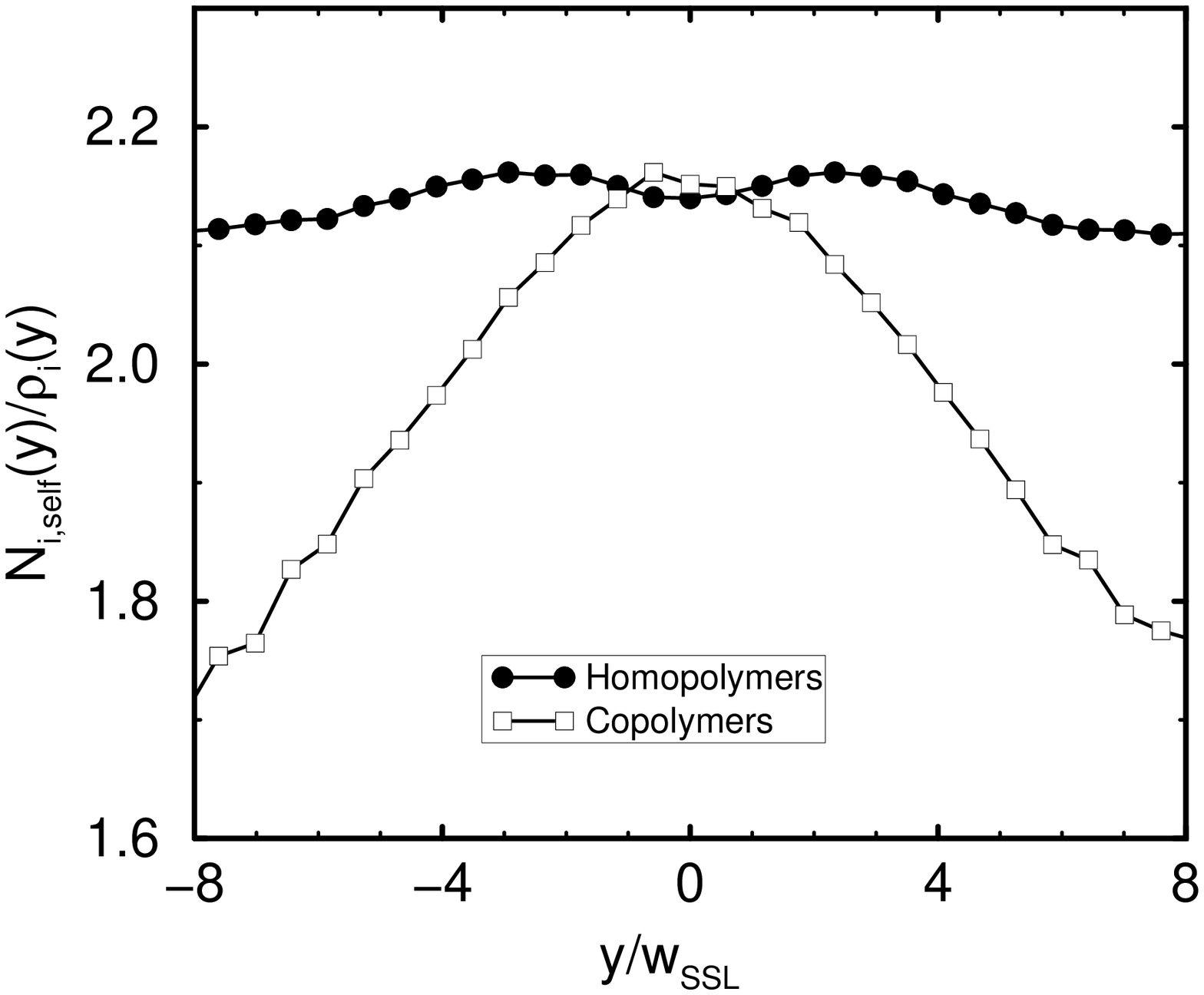,width=160mm,height=140mm}
}
\end{picture}
\vfill
\normalsize
{\tt
\noindent
Figure 8\\
Werner et al
}

\newpage
\pagestyle{empty}

\LARGE
\unitlength=1mm
\begin{picture}(150,150)
\put(-20,0){
\psfig{figure=\dir/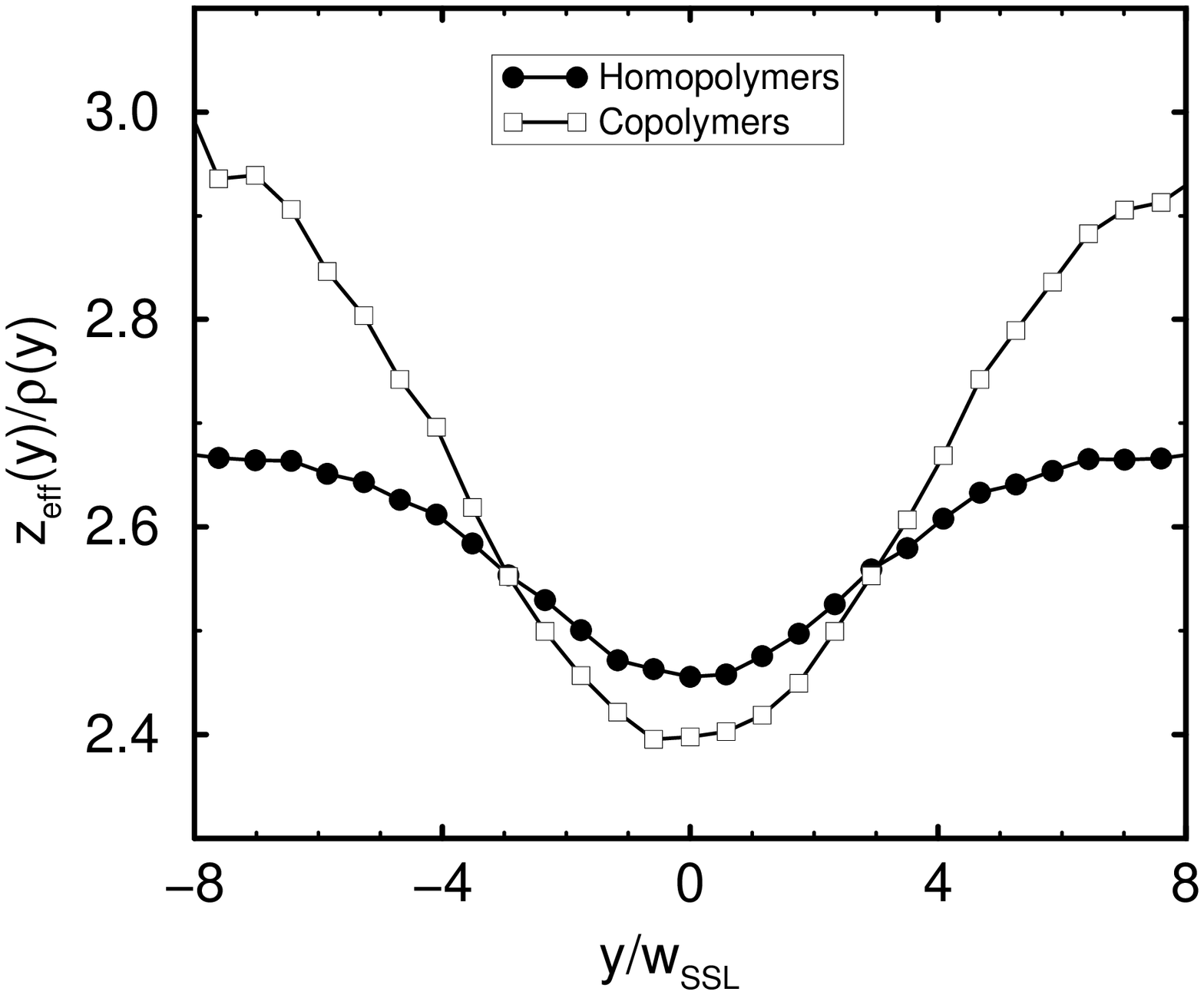,width=160mm,height=140mm}
}
\end{picture}
\vfill
\normalsize
{\tt
\noindent
Figure 9\\
Werner et al
}

\newpage
\pagestyle{empty}

\LARGE
\unitlength=1mm
\begin{picture}(150,150)
\put(-20,0){
\psfig{figure=\dir/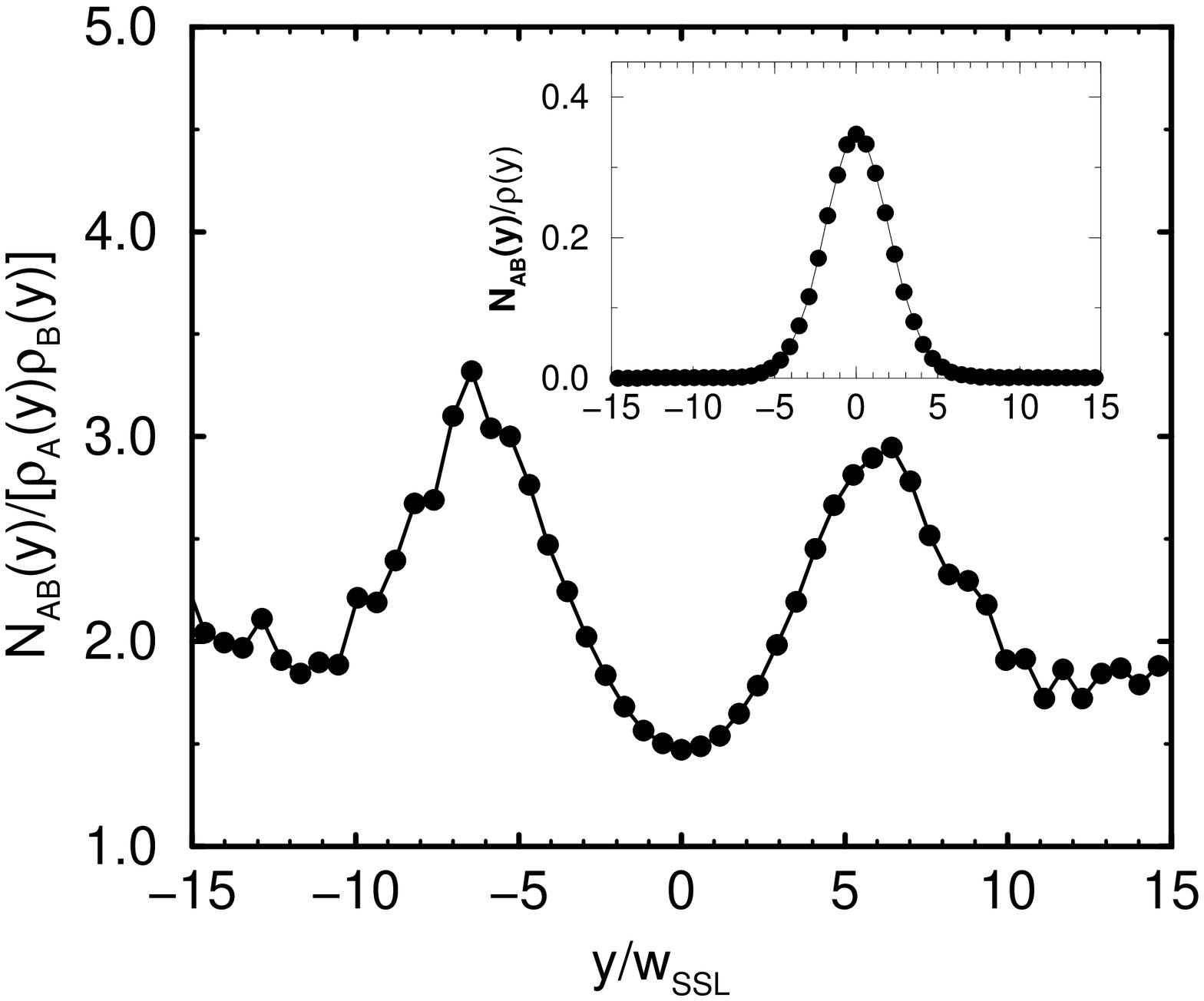,width=160mm,height=140mm}
}
\end{picture}
\vfill
\normalsize
{\tt
\noindent
Figure 10\\
Werner et al
}


\begin{thebibliography}{99}
	\bibitem{mixt} see, for example, 
Paul, D.R.; Newman, S.
{\it Polymer Blends}; 
Academic Press: New York, 1978;
\u{S}olc, K. (Ed.) 
{\it Polymer Compatibility and Incompatibility -- Principles and Practices}; 
Harwood: Chur, 1982; 
Walsh, D.S.; Higgins, J.S.; Maconnachie, A.
{\it Polymer Blends and Mixtures}; 
Martinus Nijhoff: Dordrecht, 1985;
Thomas, E.L. (Ed.) 
{\it Materials Science and Technology, Vol. 12: Structure and Properties of Polymers}; 
VCH: Weinheim, 1993.
	\bibitem{degennes} 	
De Gennes, P.-G. 
{\it Scaling Concepts in Polymer Physics};
Cornell University Press: Ithaca, NY, 1979.
	\bibitem{polint}
Sanchez, I.C. (Ed.) 
{\it Phy\-sics of Po\-ly\-mer Sur\-fa\-ces and In\-ter\-fa\-ces}; 
But\-ter\-worth--Hei\-ne\-mann: Bos\-ton, 1992;
Fleer, G.J.; Cohen Stuart, M.A.; Scheutjens, J.M.H.M.; Cosgrove, T.; Vincent, B.
{\it Polymers at Interfaces}; 
Chapman \& Hall: London, 1993.
	\bibitem{cop} 
Bates, F.S.; Fredrickson, G.H.
{\it Annu. Rev. Phys. Chem.} {\bf 1990}, {\em 41}, 525; 
{\it Ann. Rev. Mater. Sci.}, submitted.
	\bibitem{B4} 
Binder, K. 
{\it Advances in Polymer Science} {\bf 1994}, {\em 112}, 181.
%
	\bibitem{AGK} 
Anastasiadis, S.H.; Gancarz, I.; Koberstein, J.T.
{\it Macromolecules} {\bf 1989}, {\em 22}, 1449.
	\bibitem{BR} 
Brown, H.R. 
{\it Macromolecules} {\bf 1989}, {\em 22}, 2859;
Brown, H.R.; Deline, V.R.; Green, P.F.
{\it Nature} {\bf 1989}, {\em 341}, 221;
Brown, H.R.; Char, K.; Deline, V.R.; Green, P.F.
{\it Macromolecules} {\bf 1993}, {\em 26}, 4155.
	\bibitem{Dai1}
Dai, C.A. {\it et al} 
{\it Phys. Rev. Lett.} {\bf 1994}, {\em 73}, 2472.
	\bibitem{LC} 
Lee, Y.; Char, K.
{\it Macromolecules} {\bf 1994}, {\em 27}, 2603.
	\bibitem{phdexp} 
Kinning, D.J.; Winey, K.I.; Thomas, E.L.
{\it Macromolecules} {\bf 1988}, {\em 21}, 3502;
Kinning, D.J. {\it et al} 
{\it J. Chem. Phys.} {\bf 1989}, {\em 90}, 5806;
Winey, K.I.; Thomas, E.L.; Fetters, L.J.
{\it J. Chem. Phys.} {\bf 1991}, 9367; 
{\it Macromolecules} {\bf 1992}, {\em 25}, 422; 
Hashimoto, T. {\it et al}
{\it Macromolecules} {\bf 1993}, {\em 25}, 1433;
Disko, M.M. {\it et al} 
{\it Macromolecules} {\bf 1993}, {\em 26}, 2983.
	\bibitem{phdth}
Leibler, L.; Pincus, P.A. 
{\it Macromolecules} {\bf 1984}, {\em 17}, 2922;
Semenov, A.N.
{\it Macromolecules} {\bf 1993}, {\em 26}, 2273.
	\bibitem{HS} 
Holyst, R.; Schick, M.
{\it J. Chem. Phys.} {\bf 1992}, {\em 96}, 7728.
	\bibitem{MWM}
Matsen, M.W.
{\it Phys. Rev. Lett.} {\bf 1995}, {\em 74}, 4225;
{\it Macromolecules} {\bf 1995}, {\em 28}, 5765.
	\bibitem{RI}
Israels, R. {\it et al}
{\it J. Chem. Phys.} {\bf 1995}, {\em 102}, 8149.
	\bibitem{B}
Bates, F.S. {\it et al}
{\it Phys. Rev. Lett.} {\bf 1995}, {\em 75}, 4429.
	\bibitem{MS}
Schick, M.
{\it Ber. d. Bunsenges.}, in press.
	\bibitem{L2} 
Leibler, L.
{\it Makromol. Chem., Macromol. Symp.} {\bf 1988}, {\em 16}, 1. 
	\bibitem{SE} 
Semenov, A.N.
{\it Macromolecules} {\bf 1992}, {\em 25}, 4967.
%
	\bibitem{LH} 
L\"owenhaupt, B.; Hellmann, G.P.
{\it Colloid Polym. Sci.} {\bf 1990}, {\em 268}, 885.
	\bibitem{BHP} 
Bucknall, D.G.; Higgins, J.S.; Penfold, J.
{\it Physica B} {\bf 1992}, {\em 180 \& 181}, 468.
	\bibitem{SKHT} 
Shull, K.R.; Kramer, E.J.; Hadziioannou, T.; Tang, W.
{\it Macromolecules} {\bf 1990}, {\em 23}, 4780.
	\bibitem{Dai2} 
Dai, K.H.; Norton, L.J.; Kramer, E.J.
{\it Macromolecules} {\bf 1994}, {\em 27}, 1949;
Dai, K.H.; Washiyama, J.; Kramer, E.J.
{\it Marcomolecules} {\bf 1994}, {\em 27}, 4544.
	\bibitem{R1} 
Russell, T.P. {\it et al}
{\it Macromolecules} {\bf 1991}, {\em 24}, 1575.
	\bibitem{GR} 
Green, P.F.; Russell, T.P.
{\it Macromolecules} {\bf 1991}, {\em 24}, 2931.
	\bibitem{SMR} 
Shull, K.R.; Mayes, A.M.; Russell, T.P.
{\it Macromolecules} {\bf 1993}, {\em 26}, 3929.
%
	\bibitem{MA1}
Mayes, A.M.; Russell, T.P.; Satija, S.K.; Majkrzak, C.F.
{\it Macromolecules} {\bf 1992}, {\em 25}, 6523.
	\bibitem{MA2}
Mayes A.M. {\it et al} 
{\it Macromolecules} {\bf 1993}, {\em 26}, 1047.
%
	\bibitem{HF1} 
Helfand, E.; Tagami, Y. 
{\it J. Polym. Sci. B} {\bf 1971}, {\em 9}, 741; 
{\it J. Chem. Phys.} {\bf 1971}, {\em 56}, 3592;
{\bf 1972}, {\em 57}, 1812;
Helfand, E.; Sapse, A.M. 
{\it J. Chem. Phys.} {\bf 1975}, {\em 62}, 1327.
	\bibitem{HF2} 
Helfand, E. 
{\it J. Chem. Phys.} {\bf 1975}, {\em 62}, 999.
	\bibitem{FR1}
Tang, H.; Freed, K.F. 
{\it J. Chem. Phys.} {\bf 1991}, {\em 94}, 6307;
Dudowicz, J; Freed K.F. 
{\it Macromolecules} {\bf 1990}, {\em 23}, 1519;
Lifschitz, M.; Freed, K.F. 
{\it J. Chem. Phys.} {\bf 1993}, {\em 98}, 8994.
	\bibitem{FR2}
Freed, K.F.
{\it J. Chem. Phys.} {\bf 1995}, {\em 103}, 3230.
	\bibitem{NH} 
Noolandi, J.; Hong, K.M.
{\it Macromolecules} {\bf 1982}, {\em 15}, 482; 
{\bf 1984}, {\em 17}, 1531.
	\bibitem{SK} 
Shull, K.R.; Kramer, E.J.
{\it Macromolecules} {\bf 1990}, {\em 23}, 4769.
	\bibitem{S} 
Shull, K.R.
{\it Macromolecules} {\bf 1993}, {\em 26}, 2346.
%
	\bibitem{KB1} 
Binder, K.
{\it J. Chem. Phys.} {\bf 1983}, {\em 79}, 6387;
{\it Phys. Rev. A} {\bf 1984}, {\em 29}, 341.
	\bibitem{FT} 
Fischel, L.B.; Theodorou, D.N.
{\it Faraday Trans.} {\bf 1995}, {\em 91}, 2381.
%
	\bibitem{MDB}
Minchau, B.; D\"unweg, B.; Binder, K.
{\it Pol. Comm.} {\bf 1990}, {\em 31}, 348.
	\bibitem{FB} 
Fried, H.; Binder, K.
{\it J. Chem. Phys.} {\bf 1991}, {\em 94}, 8349;
{\it Europhys. Lett.} {\bf 1991}, {\em 16}, 237.
	\bibitem{BF}
Binder, K.;  Fried, H.
{\it Macromolecules} {\bf 1993}, {\em 26}, 6878.
	\bibitem{PHB}
Pan, H. {\it et al}
{\it Macromolecules} {\bf 1993}, {\em 26}, 2860.
	\bibitem{WM} 
Wang, Y.; Mattice, W.L.
{\it J. Chem. Phys.} {\bf 1993}, {\em 98}, 9881; 
Wang, Y.; Li, Y.; Mattice, W.L.
{\it J. Chem. Phys.} {\bf 1993}, {\em 99}, 4068. 
	\bibitem{PSB} 
Peng, G.; Sommer, J.-U.; Blumen, A., preprint.
%
	\bibitem{cifra}
Cifra, P., Poster presented on the {\em International Workshop on
 Wetting and Self-Organization in Thin Liquid Films}, Konstanz, September 1995.
	\bibitem{MBO} 
M\"uller, M.; Binder, K.; Oed, W.
{\it J. Chem. Soc. Faraday Trans.} {\bf 1995}, {\em 91}, 2369.
	\bibitem{SM} 
Schmid, F.; M\"uller, M.
{\it Macromolecules} {\bf 1995}, {\em 28}, 8639.
	\bibitem{B3} 
Binder, K. (Ed.)
{\it Monte Carlo and Molecular Dynamics Simulations in Polymer Science}; 
Oxford University Press: Oxford, 1995.
	\bibitem{CK} 
Carmesin, I.; Kremer, K. 
{\it Macromolecules} {\bf 1988}, {\em 21}, 2819.
	\bibitem{bfm}
Paul, W.; Binder, K.; Heermann, D.W.; Kremer, K.
{\it J. Phys. II (Paris)} {\bf 1991}, {\em 1}, 37;
Deutsch, H.-P.; Binder, K.
{\it Macromolecules} {\bf 1992}, {\em 25}, 6214;
{\it J. Phys. II (Paris)} {\bf 1993}, {\em 3}, 1049.
	\bibitem{MP} 
M\"uller, M.; Paul,  W.
{\it J. Chem. Phys.} {\bf 1993}, {\em 100}, 719.
	\bibitem{MB} 
M\"uller, M.; Binder, K.
{\it Macromolecules} {\bf 1995}, {\em 28}, 1825.
	\bibitem{SB} 
Schmid, F.; Binder, K.
{\it Phys. Rev. B} {\bf 1992}, {\em 46}, 13553.
	\bibitem{marcus}
M\"uller, M.; Schick, M.; document in preparation. 
	\bibitem{KP}  
Kratki, O.; Porod, G.
{\it Rec. Trav. Chim.} {\bf 1949}, {\em 68}, 1106;
Saito, N; Takahashi, K.; Yunoki, Y. 
{\em J. Phys. Soc. Jpn.} {\bf 1967}, {\em 22}, 219.
	\bibitem{MF} 
Morse, D.C.;  Fredrickson, G.H.
{\it Phys. Rev. Lett.} {\bf 1994}, {\em 73}, 3235.
	\bibitem{MM} 
Matsen, M.W.; Schick, M.
{\it Phys. Rev. Lett.} {\bf 1994}, {\em 72}, 2660;
Matsen, M.W.
{\it J. Chem. Phys.}, {\bf 1996}, {\em 104}, 7758.
	\bibitem{WV}
Weyersberg, A.; Vilgis, T.A.
{\it Phys. Rev. E} {\bf 1993}, {\em 48}, 377.
	\bibitem{F2} 
Schmid, F., {\it J. Chem. Phys.}, {\bf 1996}, {\em 104}, 9191.
%
\end{thebibliography}
\end{document}